\documentstyle[12pt]{article}
\begin{document}
\baselineskip=24pt
\noindent
{\large Disorder and Integral Quantum Hall Effect}

\noindent
N. Kumar\\
Raman Research Institute, C.V.Raman Avenue, Bangalore 560080, India\\

~\\

\begin{abstract}
The quantum Hall conductance of a disordered two-dimensional gas of
non-interacting electrons is re-examined for its integrity against
disorder in the limit of no mixing between different Landau levels. The
exact one-electron eigenstates of the disordered system are shown to be
current carrying, with exactly the same Hall current as in the absence of
disorder. There are no localized states. Accordingly, each extensively
degenerate Landau level, now broadened out by the disorder, continues to
contribute exactly one quantum of Hall conductance ($e^2/2\pi\hbar$). In
the absence of any localized (non-current carrying) states, the Hall
plateaus can now arise only through an actual gap in the density of states
separating the broadened Landau levels.  Implications for 2D localization
are discussed.
\end{abstract}

The Hall conductance \(\sigma_{xy}\) of a two-dimensional spin-polarized
electron gas (2DEG) is well known to be stepped in units of the quantum of
conductance \(e^2/2\pi\hbar\). This is the Integral Quantum Hall Effect (IQHE).$^{1,2}$  Such a 2D electron system obtains as an
inversion/accumulation layer in a gated Si MOSFET (metal-oxide
semiconductor field effect transistor) or a modulation doped GaAs-AlGaAs
heterostructure.$^3$ The quantized conductance steps are observed
experimentally as function of the strength of the external magnetic field
($B$) applied perpendicular to the plane of the 2DEG. The field tunes the
density-of-states in the Landau levels, and thus controls their filling
for a fixed areal number density of electrons. Alternatively, one can vary
the bias (the gate) voltage $V_g$ that tunes the areal number density of
electrons of the 2DEG and thus, for a fixed magnetic field, again controls the
filling of the Landau levels. The conductance steps are separated by
plateaus of constant conductance. While the conductance steps are of a
universal height, the plateau-widths are sample dependent. There are
gauge-theoretic global arguments for the universality of the quantized
conductance steps.$^{4,5}$  The observed accuracy of quantization of the
Hall conductance, to better than 1 part in 10$^7$, corroborates well with
the robustness of these general arguments. In order to view the present
work in the proper perspective, we have to review briefly the generally accepted
explanation for the plateaus. The plateaus are supposed to arise as a
consequence of the disorder induced Anderson localization in the imperfect
samples.$^6$  Disorder, which is ubiquitous, is expected to broaden out an
otherwise extensively degenerate Landau level over an energy interval
($\Delta$, say) centered at the unperturbed Landau level. Further, the
tail states (away from the centre) are assumed to get localized and
separated from the extended middle states by mobility edges. Now, as the
{\sl Fermi-level is swept} through these localized states, by varying the
gate voltage, say, one expects no change in the Hall current $-$ the
filling/emptying of the non-current carrying (localized) states can not
alter the total current. And hence the conductance plateaus.  The width of the
plateau then corresponds to the energy interval between the two mobility
edges separating the adjacent Landau levels.  The step, on the other hand,
corresponds to the transit of the Fermi-level across the band of extended
middle states between two mobility edges bracketing a given Landau level
$-$ gradual filling/emptying of these current carrying states changes the
total Hall current. The above widely accepted argument clearly implies
that the plateau is not due to any gap in the density of states separating
the Landau levels. It is rather due to the mobility gap $-$ the
localization of the tail states that lie between the impurity-broadened
Landau levels.  Indeed, in this picture no disorder would mean no
plateaus, and hence no experimental visualization of the IQHE-steps.$^6$
The quantization of the Hall conductance (\(e^2/2\pi\hbar\) per Landau
level) would, of course, demand then that the loss of the Hall current due
to localization of some of the states be made up exactly by a
correspondingly extra current to be carried by what extended states there
might still be there.$^6$  On the other extreme, strong disorder is also
known to wipe out the IQHE. This picture has motivated detailed
studies$^{6,7}$ of the effect of some model impurity potentials, e.g.,
delta-function, on the Hall conductance  and on the spatial distribution of the Hall current. It
has also led to detailed numerical and theoretical discussion of the
localization of eigenfunctions and of the spectral aspect of the IQHE in
the presence  of disorder.$^{8-13}$  More specifically, it has been
argued that in the limit of strong field ($B$), or weak disorder ($\Delta$),
there exists a single critical sub-extensively degenerate extended-state level
at the centre of the impurity broadened Landau level, rather than a band of
extended middle states separated from the localized tail states by
mobility edges.$^{8,9}$  In the limit  of weak field, however, these extended states are believed to {\sl float-up} energetically so as to
conform to known absence of extended states in a 2D disordered system in zero
field.$^{10-13}$ 

The present work addresses questions relating  precisely to 
these fundamental aspects. In particular, we show that (1) the exact
one-electron eigenstates of the disordered magnetized 2DEG in the QHE
geometry are all current-carrying,  (2) these exact one-electron
eigenstates carry exactly the same current as the corresponding unperturbed
eignstates do in the absence of disorder, and (3) the conductance
plateaus arise then from the actual gap in the density of states separating the
adjacent Landau levels. Our treatment of IQHE is exact in the absence of
inter-Landau level mixing due to disorder. This would correspond to the
adiabatic limit$^7$ of high magnetic field giving the Landau-level spacing
$>>$ random impurity potential fluctuation on the magnetic length scale.
Thus, except for the abrupt, or singular potentials, e.g., a
delta-function, the neglect of inter-level mixing is believed to be valid
in the high-field limit despite the corresponding enhancement of
degeneracy.$^{14}$  In any case there is no exact microscopic
theory, to the best of our knowledge, that takes into account the
inter-level mixing due to disorder. Also, we note that we are concerned
here with IQHE and not the Fractional Quantum Hall Effect (FQHE),$^{15}$
\(\sigma_{xy} = \nu e^2/2\pi\hbar\) with \(\nu\) = 1/3, 2/5, 2/3, $\dots$.
This is known to be essentially a many-body problem involving
electron-electron interaction in a fractionally filled Landau
level.$^{16}$ Some of our exact results, however, do remain valid in the
presence of interaction, again in the absence of  any inter-level 
configurational
mixing. 

Let the sample \(L_x X L_y\) be  magnetized with a uniform magnetic field \({\bf
B} = -B\hat{z}\) along the z-axis, and electrified with an
electric field along the x-axis giving a bias potential \(V(x) = -
|e|E_x\). Let the total Hall current for this QHE-geometry be \({\bf J} =
J\hat{y}\) along the y-axis. We will assume the thermodynamic limit ($L_x,
L_y \rightarrow \infty$) and zero temperature throughout. The one-electron
Hamiltonian $H$ with the vector potential \({\bf A} = (0, -Bx)\) in the
Landau gauge is then  
\begin{equation} H = H_o + H_1
\end{equation}
with the unperturbed Hamiltonian (without disorder)
\begin{equation}
H_o = - \frac{1}{2m} \left(-ih \frac{\partial}{\partial x}\right)^2 -
\frac{1}{2m} \left(-i\hbar \frac{\partial}{\partial y} - \mid e
\mid Bx\right)^2 - \mid e \mid E_x\,\,,
\end{equation}
and $H_1$ the perturbation due to the potential disorder. The
unperturbed Landau-Stark Hamiltonian has the well known eigenstates and
eigenvalues.$^6$ 
\begin{eqnarray}
\psi_{pn}^o(x,y) &=& A_n^{-1/2} exp(ipy/\hbar) exp \left[- \frac{1}{2}
\left(\frac{x-x_p}{l}\right)^2\right] \dot H_n (\frac{x-x_p}{l}) \nonumber\\
E_{np}^o &=& \hbar w_c(n + \frac{1}{2}) + (\frac{cE}{B})p\,\,,
\end{eqnarray}
where \(l = (\frac{\hbar c}{\mid e \mid B})^{1/2}\) the magnetic length,
\(w_c = \mid e \mid B/mc\) the cyclotron frequency, \(p =
\left(\frac{2\pi\hbar}{L_y}\right) k \,\,{\rm with}\,\, k = 0, \pm 1, \pm 2,
\cdots\) for periodic boundary condition in the y-direction, and \(x_p =
\frac{mc^2E}{\mid e \mid B^2} - \frac{c}{\mid e \mid B} p\).  Also, we
have the normalization constant \(A_n = (\pi^{1/2} l L_y 2^n n!)\,\,\). 
These exact eigenstates are readily seen to be current carrying. The
operator for the $y$-component of the total current (the Hall current) is
\begin{equation}
\hat{J}_y = - \frac{\mid e \mid}{L_y} \,\,\frac{1}{m} \left( -ih
\frac{\partial}{\partial y} - \frac{\mid e \mid}{c} Bx\right)
\end{equation}
We readily verify that its expectation value in the eignstate \(\psi_{np}
\equiv \mid np\big>\) is
\begin{equation} 
J_y \equiv \Big<np \mid \hat{J} \mid np \Big> = \mid e \mid c \frac{E}{BL_y}
\equiv J_o\,\,.
\end{equation}
Thus, each eignfunction carries exactly the same current $J_o$.

With these elementary results for the unperturbed system ($H_o$) in
hand, we will now show that the exact eigenstates of the disordered system
($H$), unknown as they may be, are also current carrying, and that each of
these carries exactly the same  Hall current as in the absence of disorder
($H_1$) so long as we assume no inter-Landau-Level mixing.
For this, consider the effect of disorder $H_1$ on the Landau level $n$.
Inasmuch as the exact eigenstates $\psi_{np}^o$ of $H_o$ form effectively
a complete set of functions for the Landau level ($n$), because of no
inter-level mixing, we can express the i$^{th}$ exact eigenstate, \( \psi_n^i \), as
a linear combination of the unperturbed \( \psi_{np}^o\)'s :
\begin{equation}
\mid \psi_n^i\big> = \sum_p C_{np}^i \mid \psi_{np}\Big>
\end{equation}
with the unitarity condition on the coefficients $C_{np}^i$
\begin{equation}
\sum_n C_{np}^{i*}\,\,C_{np}^i = \delta_{pp^\prime}
\end{equation}
We now consider the expectation value of \( \hat{J}_y \) in this new
exact eigenstate
\begin{eqnarray}
\Big<\psi_n^i \mid \hat{J}_y \mid \psi_n^i \Big> &=& \sum_{pp^\prime}
C_{np\prime}^{i*}\,\,C_{np} \Big<\psi_{np^\prime}^o \mid \hat{J} \mid
\psi_{np}^i\Big> \nonumber \\
&=& \sum_{pp^\prime} C_{np^\prime}^{i*}\,\,C_{np}
\delta_{pp^\prime}\frac{\mid e \mid cE}{B} \nonumber\\
&=& \frac{\mid e \mid cE}{B} \equiv J_o
\end{eqnarray}
In deriving Eq (8) we have made use of the fact that
\begin{equation}
\Big<\psi_{np} \mid \frac{\mid e \mid Bx}{c} \mid \psi_{np} \Big> =
\delta_{pp^\prime} \left(-p + \frac{mcE}{B}\right)
\end{equation}
Thus, the exact eigenstates of the disordered (electrified) Quantum Hall
system, obtained by a general unitary mixing within the sub-space of a
given Landau level, carry exactly the same total Hall current as in the
absence of disorder. This is our main observation. Clearly, the total Hall
current for the filled Landau level would then be \(\frac{J_o L_x L_y}{2\pi
l^2}\), giving Hall conductance per Landau level \( \left(\frac{J_o L_x
L_y}{2\pi l^2}\right)\Big/L_xE = \left(\frac{\mid e \mid^2}{h}\right)\), just
as for the pure system. This last result remains valid in the presence of
electron-electron interaction as well, so long as we assume no inter-Landau
level configurational mixing: A completely fixed (impurity broadened) Landau level described by
a determinantal wave function constructed from the exact $\psi_n^i$'s for
a given level $n$, clearly contributes exactly the same Hall current and
hence the same Hall conductance.

The above exact results have important consequences for the interpretation
of IQHE, for the plateaus in particular, and for the  question of
Anderson localization in 2 dimensions in the presence of a strong
perpendicular magnetic field in general. Given our result that the exact eigenstates
of the disordered system are all current carrying and that there are no
localized states, the conventional
picture of capacitative ($C$) charging ($Q = CV_g$) of the 2D interface by the
gate voltage ($V_g$) would produce a linear I-V characteristics {\sl sans}
plateaus $\,\,-\,\,$ even in the presence of spectral gaps in the density of
states. In our view the resolution of this paradox lies in the realization
that the capacitance $C$ above has been mistakenly taken to be the
geometrical capacitance. The
latter is valid when the electrodes are metallic and therefore with no gaps in
the density of states. In the presence of the spectral gap between the
(broadened) Landau levels, however, one must re-calculate the effective
interfacial capacitance from the general expression for the charging
energy that now must include the energy required to promote the electron
across the gap in the density of states. This {\sl quantum} capacitance would
then become a non-trivial function of $V_g$ itself. One expects the width
of the plateau in terms $V_g$ to reflecting the gapping of the spectrum.
(For, a sufficiently disordered system, however, when the broadened
Landau levels bridge the inter-level gaps, this would lead to no-plateaus,
in agreement with the known result that  IQHE disappears in very dirty
(strongly disordered) samples.$^{6}$  Of course, in this limit,
strictly speaking, our assumption of no inter-level mixing also becomes an
approximation.) As for the localization in a disordered 2DEG in the
presence of strong perpendicular magnetic field (and, of course, a weak
cross-electric field, the gate voltage $V_g$ that can be taken to be
arbitrarily small), our result (Eq. 8) casts serious doubts on the picture
of a (sub-extensively) degenerate, critical Landau level of extended
current carrying states bracketed by localized insulating tail states
controlling the IQHE in the strong field limit. One may have to consider
the possibility of having weakly localizable (normalizable) eigenstates
that nevertheless carry the Hall current. Alternative would be to suspect
the condition of no inter-Landau mixing no matter how strong the magnetic
field $B$ is, due presumably to the corresponding enhancement of the
degeneracy of the Landau level. Our strong-field results with no mixing,
however,   do not contradict the numerical results$^{13}$ obtained in the weak
field limit that suggest the {\sl floating-up} of the central extended
Landau states as $B \rightarrow 0$.

We would like to make a final observation in connection with the 2D
localization in a perpendicular magnetic field. It is to be noted that in 
the absence of inter-level mixing the unperturbed \(\psi_{np}^o\)'s
can be taken as a complete set of basis functions (for a given level $n$)
labeled by the quasi-continuous index $p$, giving effectively a 1D
tight-binding disordered Hamiltonian. An unusual feature of this
Hamiltonian, however, is  that the off-diagonal matrix elements are
{\sl essentially complex} random variables in that the Peierls-Landau phase
factors can not be gauged away despite the one-dimensionality. This is
because the off-diagonal matrix elements are not  {\sl
nearest-neighbouring only} on the 1D quasi continuous $p$-lattice. The diagonal
(real) matrix elements, of course, self-average to zero, given the {\sl extended} nature of \(\psi_{np}^o
(x,y)\). Such a 1D system with off-diagonal disorder calls for a detailed
study, particularly in view of the well known result for the case of
a nearest-neighbour tight-binding Hamiltonian with off-diagonal disorder,
namely that the state at the centre of the band is always extended.

In conclusion, we have shown that the exact eigenstates of a disordered
2-dimensional electron gas in a perpendicular field are all current carrying. There are no
localized states. Hall plateaus are due to spectral gaps and not the
mobility gaps. These results are exact in the adiabatic limit of no
inter-level mixing, valid for strong magnetic field, or equivalently for weak disorder.

\end{document}